# ON THE STRENGTH OF THE CARBON NANOTUBE-BASED SPACE ELEVATOR CABLE: FROM NANO- TO MEGA-MECHANICS

Nicola M. Pugno
Department of Structural Engineering, Politecnico di Torino,
Corso Duca degli Abruzzi 24, 10129, Italy; nicola.pugno@polito.it

**Abstract**

In this paper different deterministic and statistical models, based on new quantized theories proposed by the author, are presented to estimate the strength of a real, thus defective, space elevator cable. The cable, of ~100 megameters in length, is composed by carbon nanotubes, ~100 nanometers long: thus, its design involves from the nano- to the mega-mechanics. The predicted strengths are extensively compared with the experiments and the atomistic simulations on carbon nanotubes available in the literature. All these approaches unequivocally suggest that the megacable strength will be reduced by a factor at least of ~70% with respect to the theoretical nanotube strength, today (erroneously) assumed in the cable design. The reason is the unavoidable presence of defects in a so huge cable. Preliminary in silicon tensile experiments confirm the same finding. The deduced strength reduction is sufficient to pose in doubt the effective realization of the space elevator, that if built as today designed will surely break (according to the author's opinion). The mechanics of the cable is also revised and possibly damage sources discussed.

## 1. Introduction

A space elevator (Figure 1) basically consists of a cable attached to the Earth surface for carrying payloads into space (Artsutanov, 1960). If the cable is long enough, i.e., around 150Mm -a value that can be reduced by a counterweight, the centrifugal forces exceeds the gravity of the cable, that will work under tension (Pearson, 1975). The elevator would stay fixed geosynchronously. Once sent far enough, climbers would be accelerated by the Earth's rotational energy. It is clear that a space elevator would revolutionize the methodology for carrying payloads into space, and in addition at "low" cost. On the other hand, its design is very challenging.

The most critical component in the space elevator design is undoubtedly the cable, that requires a material with very high strength and low density. Considering a cable with constant section and a vanishing tension at the planet surface, the maximum stress, reached at the geosynchronous orbit (GEO), is for the Earth equal to 63GPa, even if the low carbon density (1300Kg/m$^3$) is assumed for the cable. Only recently, after the discovery of carbon nanotubes (Iijima, 1991), such a large strength has been experimentally observed (Yu et al., 2000a,b), during tensile tests of ropes composed by single walled carbon nanotubes or multiwalled carbon nanotube, expected to have an ideal strength of about 100GPa. Note that for steel (density of 7900Kg/m$^3$, strength that we assume of 5GPa) the maximum stress expected in the cable is 383GPa, whereas for kevlar (density of 1440Kg/m$^3$, strength of 3.6GPa) is 70GPa, thus much much higher than their strengths. However, a optimized cable design must consider a uniform tensile stress profile (Pearson, 1975) rather than a constant cross-section area. Accordingly, the cable could be built of any material (Pearson, 1975) by simply using a large enough "taper ratio", i.e., the maximum cross section area -at GEO- over its minimum value -at the Earth's surface. For example, for steel this value is $10^{33}$, for kevlar is $2.6 \times 10^8$ and for carbon nanotubes is only 1.9. Since the mass of the cable is proportional -as a first approximation- to the taper ratio, the feasibility of the space elevator seems to become only currently plausible thanks to the discover of carbon nanotubes (Edwards, 2000, 2003). The cable would obviously represent the largest engineering structure, hierarchically designed from the nano- (single nanotube with length of the order of a hundred of nanometers) to the mega-scale (space elevator cable with a length of the order of a hundred of megameters).

Unfortunately, the presence of even few vacancies in a single nanotube seems to play a dramatic role, as suggested by Quantized Fracture Mechanics (QFM) criteria (Pugno 2004, 2006a; Pugno and Ruoff, 2004, 2006). And in such a huge cable we expect pre-existing defects at least for statistical reasons (Carpinteri and Pugno, 2005) but also as a consequence of damage nucleation, e.g., due to micrometeorite or low-earth-orbit object impacts and atomic oxygen erosion. After a review on the mechanics of the cable, the effect on the strength of the mentioned damage typologies is considered. Accordingly, different deterministic and statistical models are presented to estimate the strength of a real, thus defective, carbon nanotube-based space elevator cable. All these methods suggest to expect a megacable strength reduced by a factor of at least ~70% with respect to the theoretical nanotube strength, corresponding to a mass increment larger than 300%. Thus, the deduced strength reduction is sufficient to pose in doubt the effective feasibility of the space elevator, that as today designed and according to the author's analysis and opinion, will undoubtedly break. Experiments and atomistic simulations, based on molecular- or quantum-mechanics on carbon nanotubes confirm our argument. Size-effects deduced by our in silicon experiments of carbon nanotube-based ropes confirm the mentioned strength reduction, in agreement with the first observations on the strength of long-meter nanotube-based ropes (Zhang et al., 2005).

Thus, the general optimism on the effective realization of the space elevator (by 15 years for $10B, see Edwards, 2000, 2003) is posed in doubt by the role of defects in the cable: as we have not been able to build a large glass cable possessing the strength of a glass whisker, perhaps we will face a similar limit during the practical realization of the space elevator cable, and for sure if the design of the cable will not dramatically be reconsidered. Accordingly, a detailed analysis on the role of defects in the cable seems to be crucial: formally, in addition to strength and density, the fracture toughness has to be taken into account and cannot be further neglected. The QFM criteria introduced by the author could help in solving, if a solution exists, the problem of a correct nanostructured megacable design, whereas classical atomistic simulations or experimental analyses remain unrealizable due to the tremendous size of the megacable.

**2. From nano- to mega-mechanics**

As mentioned, experiments and atomistic simulations cannot be performed on a so huge cable. Thus, we need a theory able to treat objects spanning from the nano- to the mega-scale. We demonstrate here that this theory must include a characteristic length, governing the considered size-scale, in contrast to the classical theories of Elasticity and in particular Linear Elastic Fracture Mechanics (LEFM; Griffith, 1921). Furthermore, LEFM has recently been generalized relaxing the hypothesis of a continuum crack propagation (Pugno 2004, 2006a; Pugno and Ruoff, 2004, 2006), introducing in a natural way a characteristic length, i.e., the "fracture quantum". In this section we apply such treatment to the smallest and to the largest object that in our planet fall down in the domain of mechanics, i.e., a nanotube, having radius of few nanometers, and the Earth itself, which has a radius of few megameters. We are going to show that a quantized theory successfully explains the deviations observed in the classical continuous counterparts, through the introduction of a fracture quantum, that varies from a fraction of a nanometer to few kilometers.

Let us consider the well-known Neuber's (1958) and Novozhilov's (1969) approach, that is the stress-analog of the energy-based QFM. It implies considering instead of the local stress, the corresponding force acting on a fracture quantum of length $a$, or equivalently the mean value of the stress $\sigma$ along it. By applying this theory for predicting the failure stress $\sigma_f$ of a nanotube with a nano-hole of radius $R$ around which a stress field $\sigma$ takes place, i.e., by setting $\langle \sigma \rangle_a = \sigma_{th}$ -where $\sigma_{th}$ is the theoretical material strength, we deduce the following failure stress $\sigma_f$:

$$\frac{\sigma_f}{\sigma_{th}} = \frac{2x^3 + 6x^2 + 6x + 2}{6x^3 + 11x^2 + 8x + 2}, \quad x = R/a \tag{1}$$

Note that according to Elasticity and by posing the maximum stress equal to the material strength, i.e., $\sigma_{max} = \sigma_{th}$, the prediction would be simply $\sigma_f/\sigma_{th} = 1/3$. In contrast, eq. (1) implies $\sigma_f/\sigma_{th} = 1/3$ only for $x \to \infty$ (large holes, eq. (1) do not consider boundary interactions), whereas for $x \to 0$, $\sigma_f/\sigma_{th} = 1$, i.e., holes with vanishing size do not affect the structural strength. A similar result is obtained by applying QFM, i.e., $\sqrt{\langle K^2 \rangle_a} = K_C$, where $K$ is the stress-intensity factor (here at the tip of a mode I crack, emanated from the hole) and $K_C$ is the fracture toughness of the material: in particular we found $\sigma_f/\sigma_{th}(x \to \infty) = 1/3.36$ and $\sigma_f/\sigma_{th}(x \to 0) = 1$. Mielke et al. (2004) and Zhang et al. (2005) performed quantum mechanical calculations using density functional theory and semiempirical methods and molecular mechanics to explore the role of vacancy defect in the fracture of carbon nanotubes under tension. Eq. (1) closely describes their predictions on strength of nanotubes containing pinhole defects, as we will discuss in details in Section 5. An example of comparison is given by the dashed line (eq. (1)) and the rhombs (atomistic simulations on a [29,29] carbon nanotube) reported in Figure 2. Instead of using the fracture quantum $a$ as a best fit parameter, we have more physically considered $a \approx 2.5 \text{Å}$, that is the distance between two adjacent chemical bonds broken during fracture: thus, the agreement is remarkable, as the deviation from the classical value of 1/3.

On the other hand, let us consider the coefficient of geostatic stress, i.e., the ratio between the horizontal and vertical geostatic stresses. The vertical stress at a depth $z$ is given by $\sigma_V = \gamma z$, where $\gamma$ is the specific weight of the Earth crust. Thus, the horizontal stress is given according to Elasticity by $\sigma_H = \nu/(1-\nu)\sigma_V$, where $\nu$ is the Poisson's ratio, and consequently the coefficient of geodetic stress becomes $K_0 = \nu/(1-\nu)$ (~0.4). In contrast, after a huge experimental analysis Brown and Hoek (1978) found the coefficient of geodetic stress in the form $K \approx K_0 + C/z$ in which $C \approx 1\text{Km}$ represents a corrective term, unexpected from classical Elasticity. Simply by considering instead of $\sigma_H$ its quantized version, i.e., $\sigma_H^* = \frac{1}{a}\int_z^{z+a}\sigma_H dz$, as a method to include the effect of the layered crust structure of the Earth, we deduce:

$$K = \frac{\sigma_H^*}{\sigma_V} = K_0 + \frac{K_0 a}{2z} \tag{2}$$

i.e., exactly the experimental relation, with $C = K_0 a/2$. Thus, in this context, $a \approx 5\text{Km}$.

Accordingly, at the nanoscale $a$ is found to be of the order of the Angström, whereas at the megascale of the kilometre. Thus, continuum theories, simply assuming $a=0$, are not appropriate in our multiscale context. This example shows that in general a simple but useful Quantized Elasticity can be formulated substituting the stress $\sigma$ with its mean value around a volume quantum $a^3$, i.e., $\sigma \to \langle\sigma\rangle_{a^3}$, recovering classical (local) Elasticity only in the limit case of $a \to 0$, and extending the Neuber's (1958) and Novozhilov's (1969) approach also for problems that do not involve a crack-propagation.

Similarly we expect very large fracture quantum in the study of geophysics, e.g., treating earthquakes as fracture instabilities in faults. In addition, the dynamic version of QFM (Pugno, 2006a) suggests the existence of a (incubation) time quantum for crack propagation, related to the time needed to generate a fracture quantum: such a time-delay has been observed of the order of microseconds in impact failures of small specimens (see Pugno 2006a), but of several hours during earthquake triggering (Gombers and Johnson, 2005), confirming our argument.

This analysis suggests that QFM is a powerful tool for studying spatial-temporal problems from the nano- to mega-scale, as required in the design of the nanotube-based space elevator cable.

## 3. Review on the Mechanics of the space elevator cable

The equilibrium between gravitational and centrifugal forces for a portion of length d$z$ ($z$=0 fixed at the Earth centre) of the space elevator cable implies (Pearson, 1975):

$$\frac{dT}{dz} = -\rho A(z)g(z), \quad g(z) = -\frac{GM}{z^2} + z\Omega^2 \qquad (3)$$

where $T$ is the tension in the cable, $A$ is its cross-section area, $G$ is the universal gravitational constant and $M$ and $\Omega$ are respectively the mass and rotational speed of the Earth. Since $T = \sigma A$, with $\sigma$ stress in the cable, two main and complementary cases can be discussed: a constant cross section area $A$, for which $dT = Ad\sigma$, or, a uniform stress cable profile, for which $dT = \sigma dA$.

Integrating eq. (3) assuming $A$=const yields:

$$\frac{\sigma(z) - \sigma_0}{\rho} = GM\left(\frac{1}{z_0} - \frac{1}{z}\right) + \frac{1}{2}\Omega^2\left(z_0^2 - z^2\right) \qquad (4)$$

Note that the term at the r.h.s of eq. (4) is only planet-dependent. Assuming $\sigma_0 = 0$ at $z_0 = R^*$ (Earth's radius) and $\rho = 1300 \, \text{Kg}/\text{m}^3$ as for carbon nanotubes, the maximum stress is reached at GEO, i.e., at $z_{GEO} = (GM/\Omega^2)^{1/3} \approx 35800 \, \text{Km}$ where the gravitational and centrifugal forces are self-balanced. For the Earth $\sigma_{max} = \sigma(z_{GEO}) \approx 63 \, \text{GPa}$ ($R^* \approx 6.38 \times 10^6 \, \text{m}$, $M \approx 5.98 \times 10^{24} \, \text{Kg}$, $G \approx 6.67 \times 10^{-11} \, \text{m}^3 \text{Kg}^{-1} \text{s}^{-2}$). Incidentally, this value corresponds to the highest strength observed in the experiments by Yu et al. (2000b) on multi-walled carbon nanotubes (MWCNT). Larger maximum stresses for more massive materials are expected to scale according to their density as described by eq. (4). Thus, only today the feasibility of the space elevator cable seems to become realistic, as a consequence of the discovery of carbon nanotubes.

By setting $T(z = R^* + L) = 0$ one derives the length $L$ of the megacable to be globally under tension, as $L \approx 150 \, \text{Mm}$ (Pearson, 1975). Note that for an hypothetical compressive load, the slenderness ($s = L/\sqrt{I/A}$, with $I$ moment of inertia) corresponding to the transition between the tensional collapse and the Euler elastic instability is $s_C = \pi\sqrt{E/(\sigma_C(1 + a/L))}$, where $E$ is the material Young modulus (e.g., 0.94TPa for a [10,10] carbon nanotube, according to the quantum mechanical calculations by Mielke at al., 2004), $\sigma_C$ is its (compressive) strength and the corrective term $a/L$ has been derived assuming Quantum Elasticity. Thus, larger sensitivity to elastic instability is expected for smaller $L/a$ and $E/\sigma_C$ ratios (nanoscale). The cable length $L$ can be reduced by a counterweight of mass $m_c$ at $z_c$, quantifiable by satisfying the equilibrium of the mass, i.e., from $\sigma(z_c)A = g(z_c)m_c$. The cable volume is $V = AL$, to which the total mass will be proportional. The cable elastic extension can be evaluated as $\Delta L = \frac{1}{E}\int_R^{R^*+L}\sigma(z)dz$.

On the other hand, integrating eq. (3) assuming $\sigma$=const yields:

$$\frac{A(z)}{A_0} = \exp\left\{\frac{\rho}{\sigma}\left[GM\left(\frac{1}{z} - \frac{1}{z_0}\right) + \frac{1}{2}\Omega^2\left(z^2 - z_0^2\right)\right]\right\} \quad (5)$$

that for $A = A_0$ at $z_0 = R^*$ gives a maximum area $A_{GEO}$ at $z_{GEO}$, for which:

$$\frac{A_{GEO}}{A_0} = \exp\left(0.776 R^* g_0 \frac{\rho}{\sigma}\right) \quad (6)$$

where $g_0 = g(z = R^*) \approx 9.78 \text{ms}^{-2}$ is the gravity acceleration at the Earth's surface. The l.h.s term in eq. (6) is the so-called taper ratio (Pearson, 1975). For example, as anticipated, for steel this value is approximately $10^{33}$, for kevlar $2.6 \times 10^8$ and for carbon nanotubes 1.9: only today the feasibility of the space elevator cable seems to become realistic. The cable length $L$ can be deduced satisfying the equilibrium of the counterweight, i.e., $\sigma A(z_c) = g(z_c) m_c$. The cable volume is $V = \int_{R}^{R^*+L} A(z) dz$, approximately proportional to $A_{GEO}/A_0$. The cable elastic extension is $\Delta L = \frac{\sigma}{E} L$.

Now let us consider the dynamics of the cable. Since for carbon nanotubes the taper ratio is small, we can assume $\partial A/\partial z \approx 0$ in the motion equation of the cable, even if tapered. The effect of the taper ratio on the longitudinal vibrations of the cable was studied in detail by Pearson (1975). Here we are going to present just a simplification of the problem, according to Edwards (2003). The transversal or longitudinal vibrations of the cable can be deduced by solving the classical equation of the motion:

$$\frac{\partial^2 u}{\partial t^2} = \frac{Y}{\rho} \frac{\partial^2 u}{\partial z^2} \quad (7)$$

where $u$ is the transversal or longitudinal displacement and $Y = \sigma$ for transversal or $Y = E$ for longitudinal oscillations. If the boundary conditions are both free or fixed, the period of the oscillations is:

$$P = \frac{2L}{q}\sqrt{\frac{\rho}{Y}} \quad (8)$$

where $q = 1, 2, 3, \ldots$ is an integer number. To avoid resonance $P$ must be far from the period of Moon (12.5h), Sun (12h) and Earth (24h). Accordingly, considering the first mode $L \neq \frac{P}{2}\sqrt{\frac{Y}{\rho}}$, and for $\rho = 1300 \text{ Kg/m}^3$, $Y = \sigma = 63 \text{GPa}$, $L \neq 157, 150, 300$ Mm. Resonance would imply transversal oscillations pumped by Moon, Sun or Earth: thus, this problem has to be considered with caution, since we are close to the realistic cable length. However, the megacable length $L$ can be modified by a counterweight, as previously described. It could also help in stabilizing the radial relative equilibrium of the megacable (Steindl and Troger, 2005).

## 4. Atomic oxygen erosion/corrosion, micrometeorite and low-earth-orbit object impacts

Damage nucleation in the cable is expected as distributed or localized, due to space debris erosion or impacts. In particular, atomic oxygen erosion will take place between 60 and 800Km, with the highest density around 100Km altitude (see Edwards, 2000, 2003). Classical theory of erosion (see Carpinteri and Pugno, 2004) assumes the material removal as proportional to the kinetic energy of the erosive particles, and consequently:

$$\frac{1}{2}v_o^2 \frac{dm_o}{dt} = k\frac{dV}{dt} \tag{9a}$$

where $v_o$ is the velocity of the atomic oxygen mass flux $dm_o/dt$ impacting on the cable volume $V$; the constant $k$ denotes the erosion resistant of the cable material. Accordingly:

$$\frac{dH}{dt} = \frac{\rho_o v_o^3}{2k} = K \tag{9b}$$

where $\rho_o$ is the atomic oxygen density and $H$ is the cable thickness.

Analogously, micrometeorite impacts, arising between 500 and 1700Km with the highest density around 1000Km altitude (see Edwards, 2000, 2003), will cause holes and/or craters in the cable. Particularly dangerous are the Leonid meteors, that transverse our solar system each 33 years, and that are expected in 2031. The Leonid debris includes dust particles and objects up to 10cm in diameter, and some debris is always permanent.
The removed volume after an impact can be estimated similarly to eqs. (9):

$$\frac{1}{2}v_m^2 m_m = k' \Delta V \tag{10a}$$

where $v_m$ is the velocity of the micrometeorite with mass $m_m$, creating a crater of volume $\Delta V$; the constant $k'$ denotes the impact resistant of the cable material. Accordingly:

$$\frac{\Delta V}{V_m} = \frac{\rho_m v_m^2}{2k'} = K' \tag{10b}$$

where $\rho_m$ is the meteorite density and $V_m$ is its volume.

Roughly speaking for large fragmentations $k \approx k' \approx \sigma_C$, where $\sigma_C$ is the macroscopic material strength (Carpinteri and Pugno, 2002); we can speculate that this estimation remains valid at all the size-scales if the corresponding size-dependent value for the structural strength $\sigma_C$ is considered (larger at the nanoscale, as a consequence of approaching the theoretical strength $\sigma_{th}$). For atomic oxygen erosion, assuming plausible values of $\sigma_C \approx 10\text{GPa}$, $\rho_o \approx 10^{-8}\,\text{Kg}/\text{m}^3$ and $v_o \approx 1\,\text{Km/s}$ we deduce $K \approx 10^{-9}\,\text{m/s}$, comparable with the experimental value of $K \approx 1\,\text{mm/month} \approx 3 \times 10^{-9}\,\text{m/s}$ (see Edwards, 2003). However, we have to note that here erosion is coupled with corrosion and thus that the process is more complex than as described. Similarly, for plausible values of $\sigma_C \approx 10\text{GPa}$, $\rho_m \approx 3000\,\text{Kg}/\text{m}^3$ and $v_m \approx 10\,\text{Km/s}$ we deduce $K' \approx 15$, comparable with the value $K' \approx 50$ suggested by Edwards (2003). Thus, for nano-fragmentation we could roughly estimate a material removal of $\Delta V \approx E_K/\sigma_{th}$ where $E_K$ denotes the kinetic energy of the projectile.

Eqs. (9) predict a steady-state erosion, whereas a catastrophic failure was experimentally deduced by treating data recorded on the MIR space station by applying a fractal theory of erosion, in which the main assumption is the substitution of the volumes in eqs. (9a) and (10a) with their fractal counterparts (i.e., the fractal domain of the energy dissipation, comprised between an Euclidean surface and volume; Carpinteri and Pugno, 2002, 2004). Thus eqs. (9) and (10) are not conservative; however a coating layer (e.g., of gold, platinum or aluminum) is expected to improve the protection of the cable against erosion/corrosion and micrometeorite impacts (Edwards, 2003).

According to eq. (10b) objects larger than ~10cm could destroy the cable. Low-earth-orbit objects (satellites and space debris larger than 10cm) is tracked by U.S. Space Command (~8000 objects). The probability of an impact of such an object on the cable is once over 250 days and could be avoided by controlling the cable position (Edwards, 2003). However, in the case of cable cut the scenario could be the following. The elastic energy per unit volume cumulated in the cable is of the order of $\psi = 1/2\,\sigma^2/E \approx 1/2(63\times10^9)^2 10^{-12} \approx 2\,\mathrm{T\,J/m^3}$. Breaking the cable will result in a pair of de-tensioning waves moving apart at the speed of $c = \sqrt{E/\rho} \approx \sqrt{10^{12}/1300} \approx 28\,\mathrm{Km/s}$. This would lead to a fragmentation of the cable, especially of the lower portion of it, that returning to Earth will encounter our atmosphere. According to the design proposed by Edwards (2000, 2003) carbon nanotube bundles of ~1cm long will work in parallel and will be connected in series by epoxy junctions; since the cable is expected 91Mm long (a counterweight will be present) and the junction will be melted due to friction with the atmosphere, the total cable is expected to be fragmented in ~$10^{10}$ segments. A terrorism attack could present the same scenario.

**5. The strength of a real, thus defective, carbon nanotube-based space elevator megacable**

In this section we present different deterministic and statistical models for predicting the strength of a real, thus defective, carbon nanotube-based space elevator cable. In addition to the previously discussed damage sources we expect unavoidable pre-existing defects in the cable simply due to statistical reasons (Carpinteri and Pugno, 2005), ultimately governed, but not controlled, by the production process. In fact, as we have not been able to build a large glass cable possessing the strength of a glass whisker, the principle of maximum likelihood ratio suggests us that we will face a similar limit during the practical realization of the space elevator cable. In other words, a defect-free huge cable is statistically unrealistic. In spite of this, it is assumed in the current design (Edwards 2000, 2003). Accordingly, we have to take into account the presence of defects to treat a real cable. Two different hypotheses of interaction between parallels nanotubes are plausible, depending on the cable construction: weak (i) or strong (ii) coupling. These two limits correspond to the two main practical realizations of the nanotube-based cable: (i) with parallel and independent nanotubes, (ii) in case forced to interact by transverse diagonal fibers or connected to form a "string" by epoxy junctions (see Edwards, 2003). The same behaviours could be obtained by a nanotube-rope designed as a macroscopic rope, i.e., without (i) or with (ii) a twisting angle (with an optimal value for nanotube load transfer around 120 degrees, see Qian et al., 2003). We are going to show that both these hypotheses and considering both deterministic (LEFM; Griffith, 1921; QFM, Pugno 2004, 2006a; Pugno and Ruoff, 2004) or statistical approaches (Weibull, 1939; Nanoscale Weibull Statistics, i.e., NWS, Pugno and Ruoff, 2006; Pugno 2004) yield the same prediction: the strength of a real space elevator cable is expected to be reduced by a factor of at least ~70% with respect to the theoretical carbon nanotube strength. Thus, is the author's opinion that, as today designed, the cable will break.

*Weak coupling* (i). This seems to be the most promising solution, as proposed by Edwards (2003). In such a case even if a nanotube breaks, it produces almost no effect on the others, due to the weak coupling between them. A crack is blocked and the chain reaction of fracture is terminated (Yakobson and Smalley, 1997). Unfortunately this positive behaviour has a negative counterpart, never mentioned in the extensive space elevator literature: just a single vacancy in a nanotube strongly affects its strength.

To demonstrate this we consider the atomistic simulations on strength of defective carbon nanotubes performed by Belytschko et al. (2002), Mielke at al. (2004) and Zhang et al. (2005) and we compare their results with the related QFM predictions. Nano-cracks of size *n* (number of adjacent atomic vacancy) or nano-holes of size *m* are considered: the index *m*=1 corresponds to the removal of an entire hexagonal ring, *m*=2 to the additional removal of the six hexagons around the former one (i.e., the adjacent perimeter of (18) atoms), *m*=3 to the additional removal of the neighboring 12 hexagonal rings (next adjacent perimeter), and so on. Quantum mechanics (QM) semi-empirical calculations (PM3 method) and Molecular Mechanics (MM) calculations (with a modified Tersoff-Brenner potential of second generation (MTB-G2) or a modified Morse potential (M)) are reported and extensively compared with the QFM predictions in Table 1. The comparison shows a relevant agreement, confirming and demonstrating that just few vacancies can dramatically reduce the strength of a single nanotube. In particular, Belytschko et al. (2002) performed atomistic molecular mechanics simulations on the fracture strength of defective nanotube containing *n* adjacent atomic vacancies, as reported in Table 1. The comparison with eq. (11) is also depicted in Figure 2 for the [80,0] nanotube (QFM (continuous line) vs. atomistic simulations (points)): in such a figure the two limit defects, a nano-hole and a nanocrack, are compared for similar sizes *n*; note the asymptotic behaviour $\sigma_f \propto n^0$ for holes and $\sigma_f \propto n^{-1/2}$ for cracks, by increasing the defect size $n = 2R/a = 2l/a$.

After having demonstrated the validity of QFM by this extensive comparison, we treat the experimental results reported by Yu et al. (2000a,b) on singlewalled carbon nanotube (SWCNT) ropes or on multiwalled carbon nanotubes (MWCNT) grown by arch discharge. Both the experiments were able to give a prediction of the fracture strength of a singlewalled carbon nanotube, assuming for the nanotube-rope/multiwalled-nanotube the load carried only by the external nanotubes/shell. The ropes were supposed to be composed by [10,10] nanotubes, thus with a diameter of 1.36nm, arranged in the closed-packed hexagonal structure, at a "contact" distance of 0.34nm. In Table 2 the experimental measurements are reported and rationalized by applying QFM (i.e., by setting $\sqrt{\langle K^2 \rangle_a} = K_C$), assuming the presence of *n* adjacent atomic vacancies, i.e., a nancocrack of length 2*l*=*na*:

$$\sigma_f = K_C \sqrt{\frac{1+\rho/2a}{\pi(l+a/2)}} = \sigma_{th} \sqrt{\frac{1+\rho/2a}{1+2l/a}}, \ \sigma_{th} = \frac{K_{IC}}{\sqrt{\pi/2a}} \quad (11)$$

where $\rho \approx a/2$ is the crack tip radius (note that LEFM, i.e., $K = K_C$, would give the trivial prediction of eq. (11) in the limit case of $\rho/a, a/l \to 0$). We have again assumed for consistency the fracture quantum *a* as coincident to the distance between two adjacent chemical bonds. Obviously, as eq. (1), eq. (11) assumes no interaction between defect and boundary. In Table 2 we have assumed values of *n* for the highest measured strength, to obtain plausible theoretical strength, that must be, as well-known, around 100GPa, see Table 1; the corresponding theoretical strength (*n*=0) is thus quantified. We note that pinhole defects seem to be more realistic than adjacent atomic vacancies, not only for chemical reasons but also as a consequence of the space debris impacts, sources of nano-holes rather than of nano-cracks. Assuming large holes ($R/a \to \infty$) and applying QFM, we predict $(\sigma_{th} - \sigma_f)/\sigma_{th} \to 70\%$. However note that in the experiments larger strength reductions, were observed suggesting the presence of more critical defects, such as elliptical holes or even truly nanocracks. Similarly, for the independent carbon nanotubes in the megacable the most plausible expectation is a strength reduction by a factor of at least ~70%.

An additional data set on MWCNT tensile experiments is today available (Barber et al., 2005), Table 3. However, the very large highest measured strength denotes an interaction between the external and internal walls, as pointed out by the same authors. Thus, the measured strength cannot be considered plausible for describing the strength of a SWCNT. Furthermore, in Table 3 we have assumed *n*=0 for the highest measure value of 259.7GPa (ideal strength), or alternatively for the case of the measured value of 109.5GPa (close to the plausible value of 100GPa). Thus, in this last case and for the higher values of strengths, sites of interactions (here treated as "negative" vacancies) between the two external layers have to be assumed; roughly speaking, the number of interaction sites can be estimated as the differences between the previous two cases, as described in Table 3; and in the context of load transfer, site of interactions are positive features. Note that we are just now going in the third-generation era of nanotensile tests (Zhu, Espinosa, 2005), suggesting that in the future rigorous experiments, by simultaneously independent stress and strain monitoring, will be possible also at the nanoscale.

The discussed tremendous defect sensitivity is confirmed by a statistical analysis based on NWS (Pugno and Ruoff, 2006). According to this theory, the probability of failure $F$ for a nearly defect-free nanotube under a tensile stress $\sigma_f$ is independent from its volume (or surface), in contrast to classical Weibull Statistics (1939), namely:

$$F = 1 - \exp\left(\frac{\sigma_f}{\sigma_0}\right)^m \qquad (12)$$

The experimental data by Yu et al. (2000a,b) are treated with NWS in Figures 3a,b. For the first data set $\sigma_0 \approx 33.9 \text{GPa}$ (SWCNT), whereas for the second one (MWCNT) a comparable value $\sigma_0 \approx 31.2 \text{GPa}$ is deduced, and for both data sets the nanoscale Weibull modulus is $m \approx 2.7$. The experiments by Barber et al. (2005) are reported in Figure 4, for which $\sigma_0 \approx 108.0 \text{GPa}$ (but not significant for the strength of a single nanotube) and $m \approx 1.8$. Note that the "nominal strength" $\sigma_0$ corresponds to a probability of failure of 63%; $\sigma_0$ is statistically found with respect to $\sigma_{th}$, reduced by a factor of about 70%, even if just few vacancies are expected to be the cause of this tremendous reduction. Note that considering a partial transfer loading between the external and internal nanotubes/shells for SWCNT ropes/MWCNT would correspond to an additional "geometrical" strength reduction. Furthermore, we have to emphasize that this definition of strength refers to a cross-section annular area of a single atomic layer thick (0.34nm) and thus the "bulk" strength (referred to the compact circular area) is expected to be scaled down proportionally to the ratio between the effective and nominal cross-section areas. However, for a single SWCNT and in this context (see eqs. (4) and (5)) this is just a matter of definition since the ratio $\sigma/\rho$ is invariant.

Thus, also for the most promising solution (i), a strength reduction by a factor of at least 70% seems to be at the moment the most plausible expectation. This is due to the strong strength reduction that just few vacancies can produce (and their presence is statistically expected): roughly speaking, the effect of a single vacancy can be deduced from eq. (1) considering $2R \approx a$, i.e., $x \approx 0.5$, for which $\sigma_f/\sigma_{th} \to 0.71$. The strength band $0.71\sigma_{th} - \sigma_{th}$ is thus forbidden, as a consequence of the crack quantization.

*Strong coupling* (ii). For such a case a single vacancy does not have this tremendous effect, as suggested by the fact that the fracture quantum will be of the order of the nanotube spacing, of the order of the nanotube diameter $d$, i.e., $a \approx d$, rather than -as in the previous case- of the order of the atomic size. Roughly speaking the effect of a vacancy can be deduced from eq. (1) considering $2R \approx d/\lambda$, i.e., $x \approx 1/(2\lambda)$, where $\lambda$ denotes the ratio between the nanotube diameter (the "characteristic size" of the microstructure in the nanotube bundle) and the atomic size (the "characteristic size" of the atomic structure in a single nanotube). We expect even larger value for $a$ than $d$ at larger size scales, as a consequence of a larger cooperation between nanotubes. Anyway, the smallest plausible value for $\lambda$ is ~10. Thus for $x \approx 1/20$, $\sigma_f/\sigma_{th}' \to 0.95$, where $\sigma_{th}' \approx \sigma_{th}/\sqrt{\lambda}$ denotes the new theoretical strength, assuming cooperation between nanotubes. And for $\sigma_{th} \approx 100 \text{GPa}$ ($\lambda \approx 10$), $\sigma_{th}' \approx 32 \text{GPa}$ whereas for $\lambda \approx 100$, $\sigma_{th}' \approx 10 \text{GPa}$: thus, a reduction by a factor of ~70% with respect to the theoretical carbon nanotube strength seems to be again unavoidable, even without defects. A vacancy will additionally reduce the strength by a factor of about 5%, as previously deduced. Summarizing, for interacting nanotubes the presence of a defect is less critical but the ideal strength is intrinsically reduced, as synthetically described by:

$$\frac{\sigma_f}{\sigma'_{th}} \approx \sqrt{\frac{1+\rho/2d}{1+2l/d}}, \quad \sigma'_{th} \approx \frac{\sigma_{th}}{\sqrt{\lambda}} \qquad (13)$$

Analogously to eq. (11) also eq. (12) must be rewritten according to the coupling, namely:

$$F = 1 - \exp N_x^\alpha N_y^\beta N_z^\gamma \left(\frac{\sigma_f}{\sigma_0}\right)^m \qquad (14)$$

where, $N_x$, $N_y$, and $N_z$ are the number of nanotubes along $x$, $y$, $z$ (longitudinal axis) respectively and $\alpha, \beta, \gamma$ are the corresponding scaling exponents. Weibull Statistics (1939) basically assumes $\alpha = \beta = \gamma = 1$, with $N = N_x N_y N_z = V/V_0$, where $V_0$ is a characteristic volume, here assumed for consistency to eq. (12) as the volume of a single nanotube (and $V$ is the megacable volume). The constants $\sigma_0$ and $m$ are in general different from those appearing in eq. (12). According to eq. (14), a size-effect for the nominal strength is predicted as:

$$\sigma_f = \sigma_0 N_x^{-\alpha/m} N_y^{-\beta/m} N_z^{-\gamma/m} \qquad (15a)$$

The previous equation is simplified if one assumes Weibull Statistics (1939), as:

$$\sigma_f = \sigma_0 N^{-1/m} \qquad (15b)$$

For consistency with the previously treated case of $N=1$, $\sigma_0$ in eqs. (12) and (14) or (15) must be the same. Eq. (15b) is the simplest scaling law for a bundle composed by $N$ nanotubes, each of them with (nominal) strength $\sigma_0$. The real problem is the determination of the three exponents in eq. (15a) for the huge space elevator cable, or for simplicity the determination of $m$ in eq. (15b).

The experimental derivation of $m$ is very complex. However, recently Zhang et al. (2005) have been able to build the first meter-long cable based on carbon nanotubes. For such a nanostructured macroscopic cable a strength over density ratio of $\sigma/\rho \approx 120-144 \,\text{KPa}/(\text{Kg}/\text{m}^3)$ was measured, dividing the breaking tensile force by the mass per unit length of the cable (the cross-section geometry was not of clear identification). The cable density was estimated to be $\rho \approx 1.5 \,\text{Kg}/\text{m}^3$, thus resulting in a cable strength of $\sigma \approx 200\text{KPa}$. Thus, we estimate for the single nanotube contained in such a cable $\sigma_f \approx 170\text{MPa}$ (carbon density of $1300\text{Kg}/\text{m}^3$), much much lower than its theoretical or measured nanoscale strength, as we expected according to the scaling of eqs. (15). For such a case, assuming the nanotubes investigated at the nanoscale to be one micron long and the cable one meter in length, from eq. (15b) we deduce $m \approx -\ln(1/10^{-6})/\ln(31/0.17) \approx 2.7$. We think that this value is only eventfully coincident with that deduced by fitting the nano-tensile experiments (that did not reveal size-effects) with NWS; we expect a larger value as soon as the proposed experimental technique will be improved for producing higher quality cables. Or also, that the power-law in eqs. (15) is a too simplified approximation; in fact applying eq. (15b) with $m=2.7$ will result in a vanishing megacable strength. Note that a densified cable with a larger value of $\sigma/\rho \approx 465 \,\text{KPa}/(\text{Kg}/\text{m}^3)$ was also realized, demonstrating the possibility of improving the technique (corresponding to $m \approx 3.3$). Now let us assume to apply the same eq. (15b) to the results on force vs. number of layers reported by Zhang et al. (2005), just to have an idea about the scaling that we have to expect by varying the number of sheets in the megacable: since for 2 layers a breaking force of ~40 mN was required, whereas for 12 layers a force of ~235 mN was measured, and a linear dependence from the other tested cases of 4, 6, 8 and 10 layers was observed, we deduce $m \approx \ln(12)/\ln((12 \times 40)/(2 \times 235)) \approx 118$, thus larger as expected. Rougly, considering this value in the previous context, nothing that the megacable volume is of the order of $10^8 \times 0.1 \times 10^{-6} = 10 \, \text{m}^3$ and a nanotube has a volume of the order of $10^{-8} \times 10^{-8} \times 10^{-6} = 10^{-22} \, \text{m}^3$, a number of $N \approx 10^{23}$ nanotubes is expected, corresponding to a megacable strength of $\sigma_f \approx 34 \times (10^{23})^{-1/118} \approx 22 \text{GPa}$.

## 6. In silicon experiments on the strength of the space elevator cable: the SE$^3$ code

The SE$^3$ code has *ad hoc* been developed for the in Silicon Experiments of the Space Elevator cable and the related Size-Effects, especially on strength. This code mainly gives as outputs the strength prediction and the damage space-time localization of the megarope. The stochastic inputs are the NWS describing the experimental strengths of the carbon nano-ropes/tubes by Yu et al. (2000a,b), i.e., $F \approx 1 - \exp\left(\frac{\sigma_f [\text{GPa}]}{34}\right)^{2.7}$ and the nanotube Young modulus $E \approx 0.94 \text{TPa}$ deduced according to quantum mechanical simulations (density functional theory) by Mielke et al. (2004).

Density functional theory simulations are based on the numerical solution of the Schrödinger equation, as well as molecular mechanics or dynamics solve the Newton equation, deriving the generalized force from a given potential. Treating single particles or atoms such methods are intrinsically limited in solving objects at the atomic- or nano-scale. On the contrary the SE$^3$ code is based on the global energy balance. The space elevator cable is assumed to be composed by weak coupled (mean field solution) stochastic-linear-elastic aligned SWCNTs (or ropes): basically a networks of stochastic springs. Thus, the role played by a particle or an atom in the atomistic simulations is here played by an entire SWCNT (or rope), and thus the size limitation is correspondingly reduced. Imagine a virtual tensile experiment on a tapered space elevator cable: the uniform stress is increased in the cable, as in the tensile test of a cable with constant cross-section area. Assuming a cable compliance $C$ and stiffness $S = C^{-1}$, the total potential energy of the system is for *T*-tension or *δ*-displacement controls respectively (*T=Sδ*):

$$W = \frac{1}{2} S \delta^2 - T\delta = -\frac{1}{2} C T^2, \quad \text{or} \quad W = \frac{1}{2} S \delta^2 \qquad (17)$$

The failure of the nanotube *j* (1<*j*<*N*) will take place when the stress acting on it, $\sigma_j$, will reach the intrinsic nanotube strength $\sigma_{fj}$, stochastically distributed according to the failure probability *F* (fitted to carbon nanotubes nanotensile tests). The energy balance during failure implies:

$$\Delta W_j + \Delta E_j + \Delta \Omega_j = 0 \qquad (18)$$

where $\Delta E_j$ is the kinetic energy released and $\Delta \Omega_j$ is the dissipated energy (due to nanotube fracture); $\Delta \Omega_j = G_f \Delta A$ where $G_f$ is the energy dissipated per unit area and $\Delta A$ is the nanotube cross-section area; $\Delta W_j = -\frac{1}{2} T^2 \Delta C$, or $\Delta W_j = \frac{1}{2} \delta^2 \Delta S$ for tension- or displacement-controls respectively; $\Delta S$ and $\Delta C$ are the global variations imposed by the breakage of the nanotube *j* (of trivial evaluation, left up to the reader). Accordingly, from the elastic energy $\Phi_j = \frac{1}{2} \frac{\sigma_{fj}^2}{E} \Delta A l^*$ stored in the nanotube (of length $l^*$) at fracture, the released kinetic, dissipated and stored energies, as well as the external work can be easily computed. Space-time damage monitoring, stress-strain curve and related size-effects can be accordingly deduced. An example of outputs for a two dimensional simulation of a 100×100 network of nanotubes is reported in Figures 5, assuming a displacement control linearly varying in time (t). Rather than a power law the numerical results suggest the validity of the size/shape scaling law proposed by Pugno (2006b):

$$\frac{\sigma_f(S/V)}{\sigma_{nano}} = \left( \frac{(\sigma_{nano}/\sigma_{mega})^2 - 1}{\ell S/V + 1} + 1 \right)^{-1/2} \quad (19)$$

giving the failure strength $\sigma_f$ for a structure of volume $V$ and surface $S$ with a nano-strength $\sigma_{nano}$ and a mega-strength $\sigma_{mega}$, where $\ell$ is a characteristic length. Note that such a scaling for the case of self-similar structures of size $L$ ($L \propto \sqrt{S} \propto \sqrt[3]{V}$) having $\sigma_{nano}/\sigma_{nano} \to \infty$ agrees with the well-known Carpinteri's scaling law (Carpinteri, 1982). Preliminary results by using the SE$^3$ code with $\sigma_{nano} = \sigma_0 = 34 \text{GPa}$ are fitted with $\sigma_{mega} \approx 15 \text{GPa}$ (and $\ell << L$). Thus again, the megarope is expected with a strength significantly reduced with respect to the ideal strength of a single nanotube.

## 7. Conclusions

Our results are based on both deterministic and statistical treatments, considering or not interaction between the nanotubes in the megacable. For the last case (current proposal) the maximum strength is predicted to be larger, but with extremely high defect sensitivity; on the contrary, for the second case the situation is opposite. In any case the strength of a real, thus defective, carbon nanotube based space elevator megacable is expected to be reduced by a factor of at least ~70% with respect to the theoretical strength of a carbon nanotube, assumed in the current design. Such a reduction is sufficient to pose in doubt the effective realization of the space elevator. Is the author's opinion that the cable, if realized as today designed, will break.

# TABLE CAPTIONS

Table 1: Atomistic simulations (Belytschko et al., 2002; Mielke at al., 2004; Zhang et al., 2005) and QFM (Pugno and Ruoff, 2004) predictions for nano-cracks of size $n$ (number of adjacent atomic vacancy) or nano-holes of size $m$. The index $m=1$ corresponds to the removal of an entire hexagonal ring, $m=2$ corresponds to the additional removal of the six hexagons around the former one (i.e., the adjacent perimeter of (18) atoms), $m=3$ considers in addition the removal of the neighboring 12 hexagonal rings (next adjacent perimeter), and so on. Quantum mechanics (QM) semi-empirical calculations (PM3 method) and Molecular Mechanics (MM) calculations (modified Tersoff-Brenner potential of second generation (MTB-G2), or modified Morse potential (M)). The symbol (+H) means that the defect was saturated with hydrogen. Symmetric and asymmetric bond reconstructions were also considered (see Mielke at al., 2004; Zhang et al., 2005 for details). The tubes are "short", if not differently specified: note that for long tubes a reduction in the strength is always observed, as an intrinsic size-effect. For nested nanotube an increment in the strength of ~5GPa is here assumed to roughly take into account the van der Walls (vdW) interaction between the walls.

Table 2: QFM (Pugno and Ruoff, 2004) applied to SWCNT assuming the presence of nano-cracks, i.e., $n$ adjacent atomic vacancies: fracture strength extracted from singlewalled carbon nanotube (SWCNT, Yu et al., 2000a) ropes or multiwalled carbon nantubes (MWCNT, Yu et al., 2000b) nano tensile tests. Note that the case corresponding to the prediction of an ideal strength of 80.6GPa (too small) is unlikely.

Table 3: QFM (Pugno and Ruoff, 2004) applied to MWCNT: experiments on fracture strength extracted from MWCNT nano-tensile tests (Barber et al., 2005). Note the estimations of the interaction sites, treated as "negative" vacancies (the difference between the columns 3 and 4 for lines 18-26 is always equal to 6, i.e., the number of vacancies that must be assumed in correspondence of a plausible ideal strength, starting from the wrong assumption of an ideal strength coincident with the highest measured value).

# TABLES

| Nanotube type | Nanocrack ($n$) and nanoholes ($m$) sizes | Strength [GPa] by QM (MTB-G2) and MM (PM3; M) atomistic or QFM calculations |
|---|---|---|
| [5,5] | Defect-free | 105 (MTB-G2); 135 (PM3) |
| [5,5] | $n$=1 (sym.+H) | 85 (MTB-G2); 106 (PM3) |
| [5,5] | $n$=1 (Asym. +H) | 71 (MTB-G2); 99 (PM3) |
| [5,5] | $n$=1 (Asym.) | 70 (MTB-G2); 100 (PM3) |
| [5,5] | $n$=2 (Sym.) | 71 (MTB-G2); 105 (PM3) |
| [5,5] | $n$=2 (Asym.) | 73 (MTB-G2); 111 (PM3) |
| [5,5] | $m$=1 (+H) | 70 (MTB-G2), 68 for long tube; 101 (PM3) |
| [5,5] | $m$=1-2 (+H) | 50 (MTB-G2), 47 for long tube; 76 (MTB-G2) |
| [5,5] | $m$=2 (+H) | 53 (MTB-G2), 50 for long tube; 78 (PM3) |
| [5,5] | Stone-Wales | 89 (MTB-G2), 88 for long tube; 125 (PM3) |
| [10,10] | Defect-free | 88 (MTB-G2); 124 (PM3) |
| [10,10] | $n$=1 (sym.+H) | 65 (MTB-G2) |
| [10,10] | $n$=1 (Asym. +H) | 68 (MTB-G2) |
| [10,10] | $n$=1 (Sym.) | 65 (MTB-G2); 101 (PM3) |
| [10,10] | $n$=2 (Sym.) | 64 (MTB-G2); 107 (PM3) |
| [10,10] | $n$=2 (Asym.) | 65 (MTB-G2); 92 (PM3) |
| [10,10] | $m$=1 (+H) | 56 (MTB-G2), 52 for long tube; 89 (PM3) |
| [10,10] | $m$=1-2 (+H) | 56 (MTB-G2), 46 for long tube; 84 (PM3) |
| [10,10] | $m$=2 (+H) | 42 (MTB-G2), 36 for long tube; 67 (PM3) |
| [50,0] | Defect-free | 89 (MTB-G2) |
| [50,0] | $m$=1 (+H) | 58 (MTB-G2); 60 (QFM) |
| [50,0] | $m$=2 (+H) | 46 (MTB-G2); 43 (QFM) |
| [50,0] | $m$=3 (+H) | 40 (MTB-G2); 37 (QFM) |
| [50,0] | $m$=4 (+H) | 36 (MTB-G2); 35 (QFM) |
| [50,0] | $m$=5 (+H) | 33 (MTB-G2); 33 (QFM) |
| [50,0] | $m$=6 (+H) | 31 (MTB-G2); 32 (QFM) |
| [100,0] | Defect-free | 89 (MTB-G2) |
| [100,0] | $m$=1 (+H) | 58 (MTB-G2); 60 (QFM) |
| [100,0] | $m$=2 (+H) | 47 (MTB-G2); 43 (QFM) |
| [100,0] | $m$=3 (+H) | 42 (MTB-G2); 37 (QFM) |
| [100,0] | $m$=4 (+H) | 39 (MTB-G2); 35 (QFM) |
| [100,0] | $m$=5 (+H) | 37 (MTB-G2); 33 (QFM) |
| [100,0] | $m$=6 (+H) | 35 (MTB-G2); 32 (QFM) |
| [29,29] | Defect-free | 101 (MTB-G2) |
| [29,29] | $m$=1 (+H) | 77 (MTB-G2); 67 (QFM) |
| [29,29] | $m$=2 (+H) | 62 (MTB-G2); 48 (QFM) |
| [29,29] | $m$=3 (+H) | 54 (MTB-G2); 42 (QFM) |
| [29,29] | $m$=4 (+H) | 48 (MTB-G2); 39 (QFM) |
| [29,29] | $m$=5 (+H) | 45 (MTB-G2); 37 (QFM) |
| [29,29] | $m$=6 (+H) | 42 (MTB-G2); 36 (QFM) |
| [47,5] | Defect-free | 89 (MTB-G2) |
| [47,5] | $m$=1 (+H) | 57 (MTB-G2); 61 (QFM) |
| [44,10] | Defect-free | 89 (MTB-G2) |
| [44,10] | $m$=1 (+H) | 58 (MTB-G2); 61 (QFM) |
| [40,16] | Defect-free | 92 (MTB-G2) |
| [40,16] | $m$=1 (+H) | 59 (MTB-G2); 63 (QFM) |
| [36,21] | Defect-free | 96 (MTB-G2) |
| [36,21] | $m$=1 (+H) | 63 (MTB-G2); 65 (QFM) |
| [33,24] | Defect-free | 99 (MTB-G2) |
| [33,24] | $m$=1 (+H) | 67 (MTB-G2); 67 (QFM) |
| [80, 0] | Defect-free | 93 (M) |
| [80, 0] | $n$=2 | 64 (M); 64 (QFM) |
| [80, 0] | $n$=4 | 50 (M); 50 (QFM) |
| [80, 0] | $n$=6 | 42 (M); 42 (QFM) |
| [80, 0] | $n$=8 | 37 (M); 37 (QFM) |
| [40, 0] (nested by a [32, 0]) | Defect-free | 99 (M) |
| [40, 0] (nested by a [32, 0]) | $n$=2 | 73 (M); 73 (QFM + vdW interaction ~5GPa) |
| [40, 0] (nested by a [32, 0]) | $n$=4 | 57 (M); 58 (QFM + vdW interaction ~5GPa) |
| [40, 0] (nested by a [32, 0]) | $n$=6 | 50 (M); 50 (QFM + vdW interaction ~5GPa) |
| [40, 0] (nested by a [32, 0]) | $n$=8 | 44 (M); 44 (QFM + vdW interaction ~5GPa) |
| [100,0] | $n$=4 | 50 (M) |
| [40,40] | $n$=4 | 54 (M) |

Table 2

| | Strength [GPa] SWCNT ropes (nano tensile tests) | Number $n$ of atomic vacancies (QFM) | | | Strength [GPa] MWCNT (nano tensile tests) | Number $n$ of atomic vacancies (QFM) | |
|---|---|---|---|---|---|---|---|
| **1** | 13 | 79 | 63 | 47 | 11 | 97 | 130 |
| **2** | 15 | 59 | 47 | 35 | 12 | 82 | 109 |
| **3** | 16 | 52 | 41 | 31 | 18 | 36 | 48 |
| **4** | 17 | 46 | 36 | 27 | 18 | 36 | 48 |
| **5** | 22 | 27 | 21 | 16 | 19 | 32 | 43 |
| **6** | 23 | 25 | 19 | 14 | 20 | 29 | 39 |
| **7** | 25 | 21 | 16 | 12 | 20 | 29 | 39 |
| **8** | 29 | 15 | 12 | 9 | 21 | 26 | 35 |
| **9** | 32 | 12 | 10 | 7 | 24 | 20 | 27 |
| **10** | 33 | 11 | 9 | 6 | 24 | 20 | 27 |
| **11** | 37 | 9 | 7 | 5 | 26 | 17 | 22 |
| **12** | 43 | 6 | 5 | 3 | 28 | 14 | 19 |
| **13** | 45 | 6 | 4 | 3 | 34 | 9 | 13 |
| **14** | 48 | 5 | 4 | 3 | 35 | 9 | 12 |
| **15** | 52 | **4** | **3** | **2** | 37 | 8 | 11 |
| **16** | | | | | 37 | 8 | 11 |
| **17** | | | | | 39 | 7 | 9 |
| **18** | | | | | 43 | 5 | 8 |
| **19** | | | | | **63** | **2** | **3** |
| **Predicted Ideal Strength [GPa]** | | *104.0* | *93.0* | *80.6 (unlikely)* | | *97.6* | *112.7* |

Table 2

| | Strength [GPa] (nano tensile tests) | Number $n$ of atomic vacancies (QFM) | |
|---|---|---|---|
| **1** | 17.4 | 277 | 49 |
| **2** | 22.3 | 169 | 29 |
| **3** | 23.7 | 149 | 26 |
| **4** | 30.0 | 93 | 16 |
| **5** | 44.2 | 42 | 7 |
| **6** | 49.3 | 34 | 5 |
| **7** | 52.7 | 29 | 4 |
| **8** | 54.8 | 27 | 4 |
| **9** | 62.1 | 21 | 3 |
| **10** | 66.2 | 18 | 2 |
| **11** | 84.9 | 11 | 1 |
| **12** | 90.1 | 9 | 1 |
| **13** | 90.3 | 9 | 1 |
| **14** | 91.1 | 9 | 1 |
| **15** | 99.5 | 8 | 1 |
| **16** | 101.6 | 7 | 0 |
| **17** | 108.5 | 6 | 0 |
| **18** | **109.5** | 6 | **0** |
| **19** | 119.1 | 5 | *-1 (interaction)* |
| **20** | 127.0 | 4 | *-2 "* |
| **21** | 132.9 | 4 | *-2 "* |
| **22** | 140.8 | 3 | *-3 "* |
| **23** | 141.0 | 3 | *-3 "* |
| **24** | 175.0 | 2 | *-4 "* |
| **25** | 231.8 | 1 | *-5 "* |
| **26** | **259.7** | **0** | *-6 "* |
| **Predicted Ideal Strength [GPa]** | | *(259.7) (unrealistic)* | *109.5* |

Table 3

# FIGURE CAPTIONS

Figure 1: The cover of the American Scientist magazine (July-August, 1997) reporting an artistic conception of the space elevator (left); and its structural scheme (right, downloaded from Wikipedia - the free encyclopedia).

Figure 2: Strength of defective SWCNT vs. hole size defined as $n = 2R/a$ (QFM dashed line, rhombs atomistic simulations), or vs. crack length $n = 2l/a$ (QFM continuous line, points atomistic simulations); the fracture quantum $a \approx 2.5 \text{Å}$ is fixed identical to the distance between adjacent broken chemical bonds, thus not as a best fit parameter.

Figure 3: NWS (Pugno and Ruoff, 2006) applied to SWCNT: experiments on fracture strength extracted from nano-tensile tests on SWCNT ropes (a; Yu et al., 2000a) and on MWCNT (b; Yu et al., 2000b) grown by arch discharge.

Figure 4: NWS applied to MWCNT: experiments on fracture strength extracted by nano-tensile tests on MWCNT grown by chemical vapour deposition (Barber et al., 2005).

Figure 5a: Broken nanotubes after 1% of elongation (e.g., 100×100 nanotube bundle).
Figure 5b: Broken nanotubes at failure and damage localization (e.g., 100×100 nanotube bundle).
Figure 5c: Stress [GPa] vs. Strain for a nanotube bundle (e.g., 100×100 nanotube bundle).
Figure 5d: External work [Nm] vs time [s] (e.g., 100×100 nanotube bundle).
Figure 5e: Stored elastic energy [Nm] vs. time [s] (e.g., 100×100 nanotube bundle).
Figure 5f: Dissipated energy [Nm] vs. time [s] (e.g., 100×100 nanotube bundle).
Figure 5g: Kinetic energy [Nm] emitted vs. time [s] (e.g., 100×100 nanotube bundle).
Figure 5h: Number of broken nanotubes vs. time during the tensile test (e.g., 100×100 nanotube bundle).

**FIGURES**

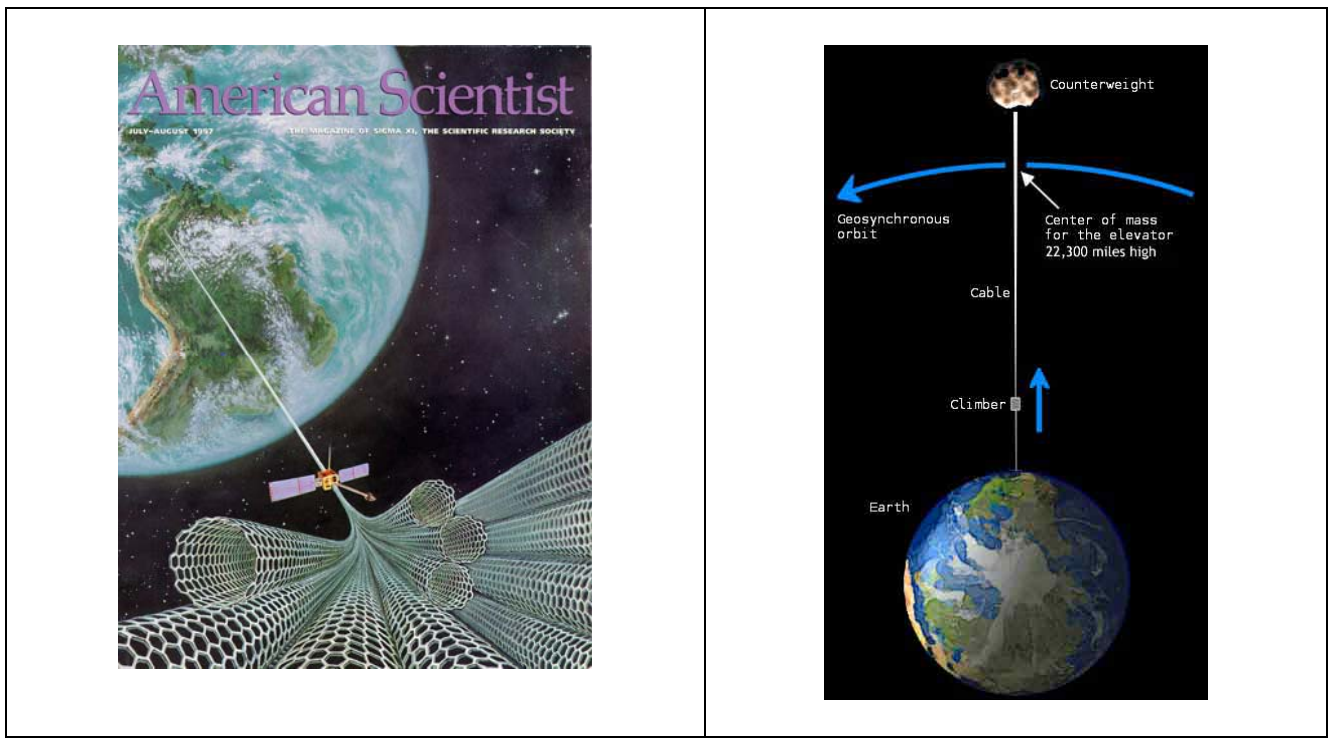

Figure 1

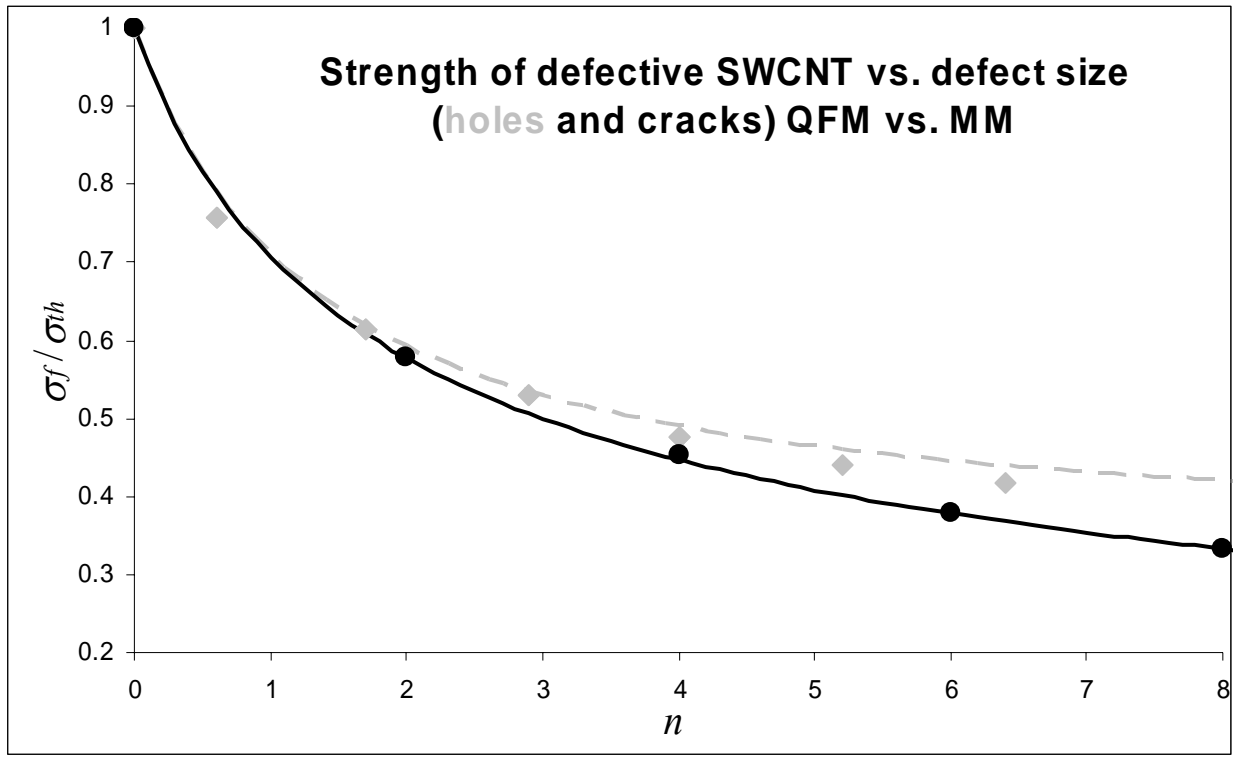

Figure 2

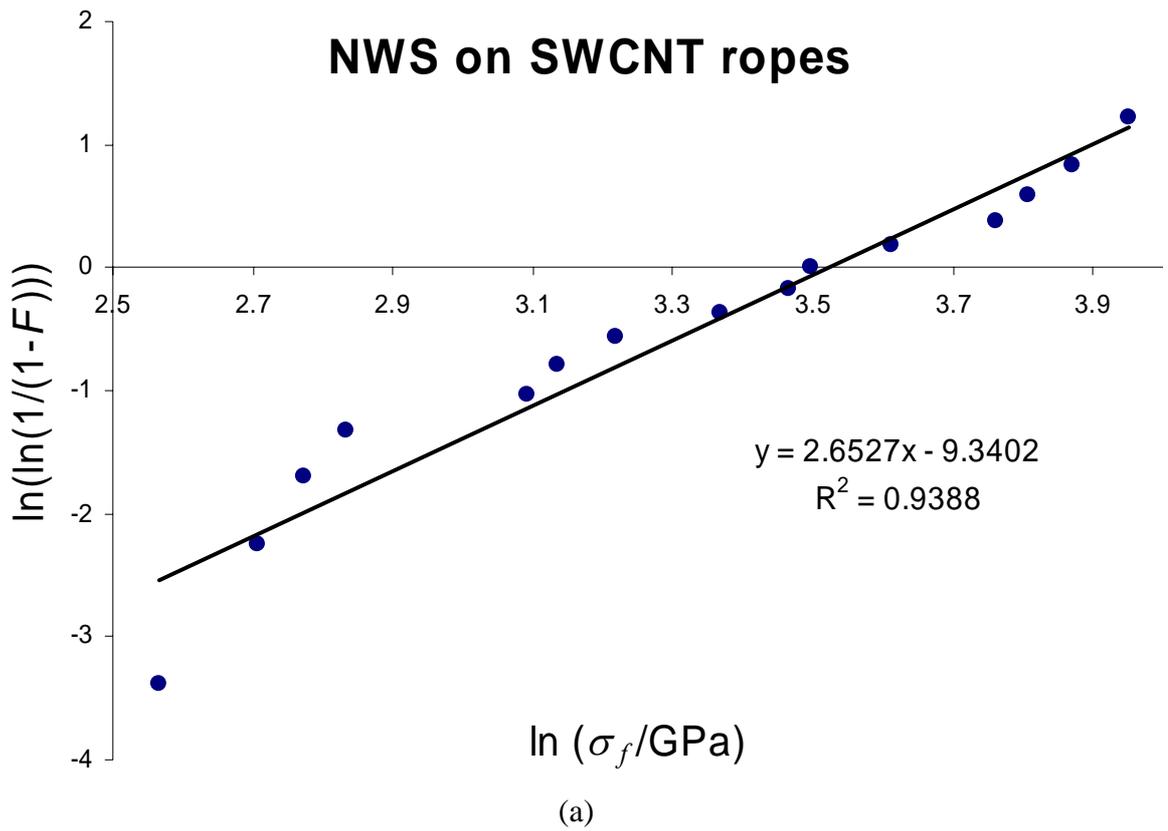

(a)

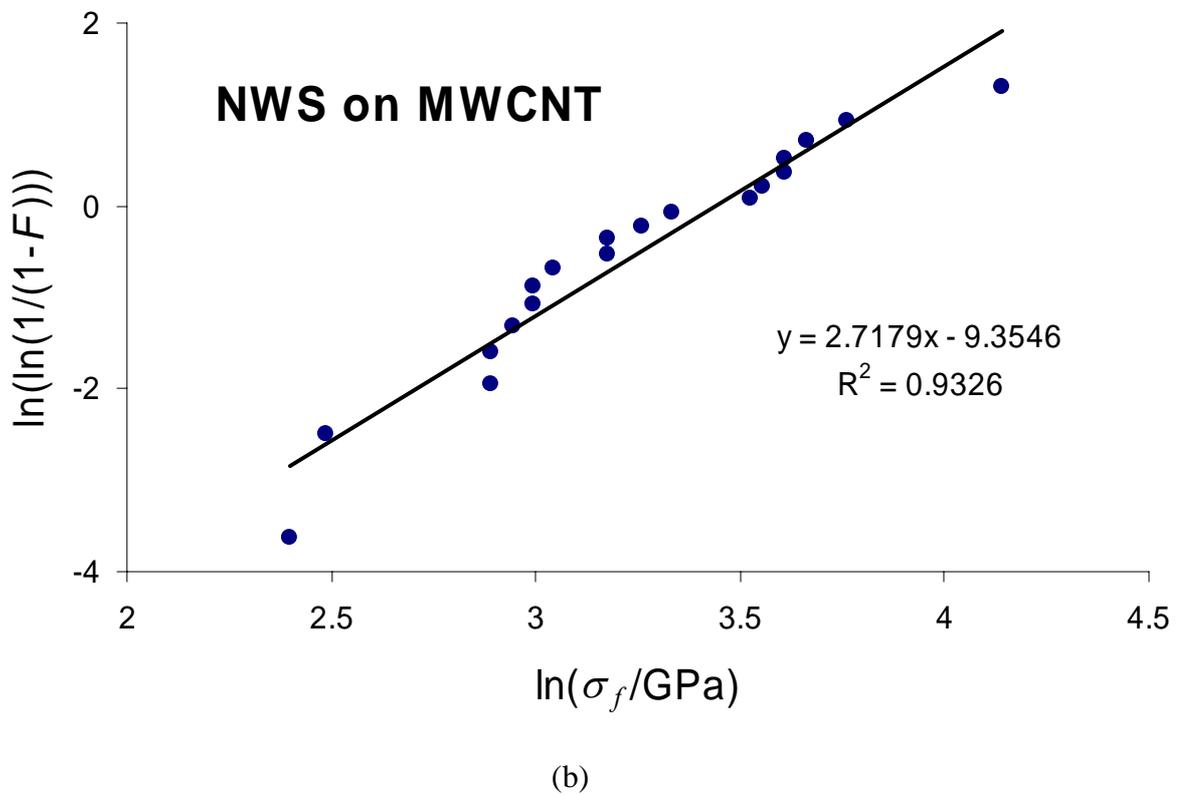

(b)

Figure 3

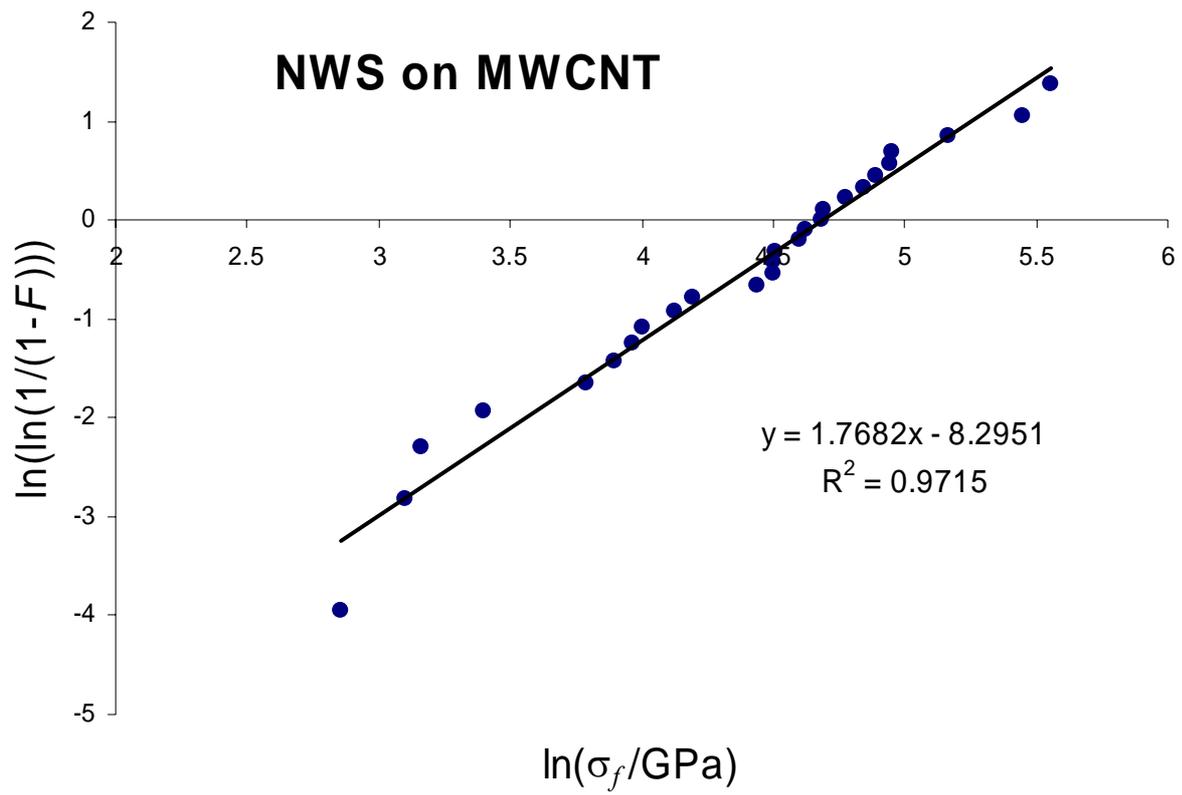

Figure 4

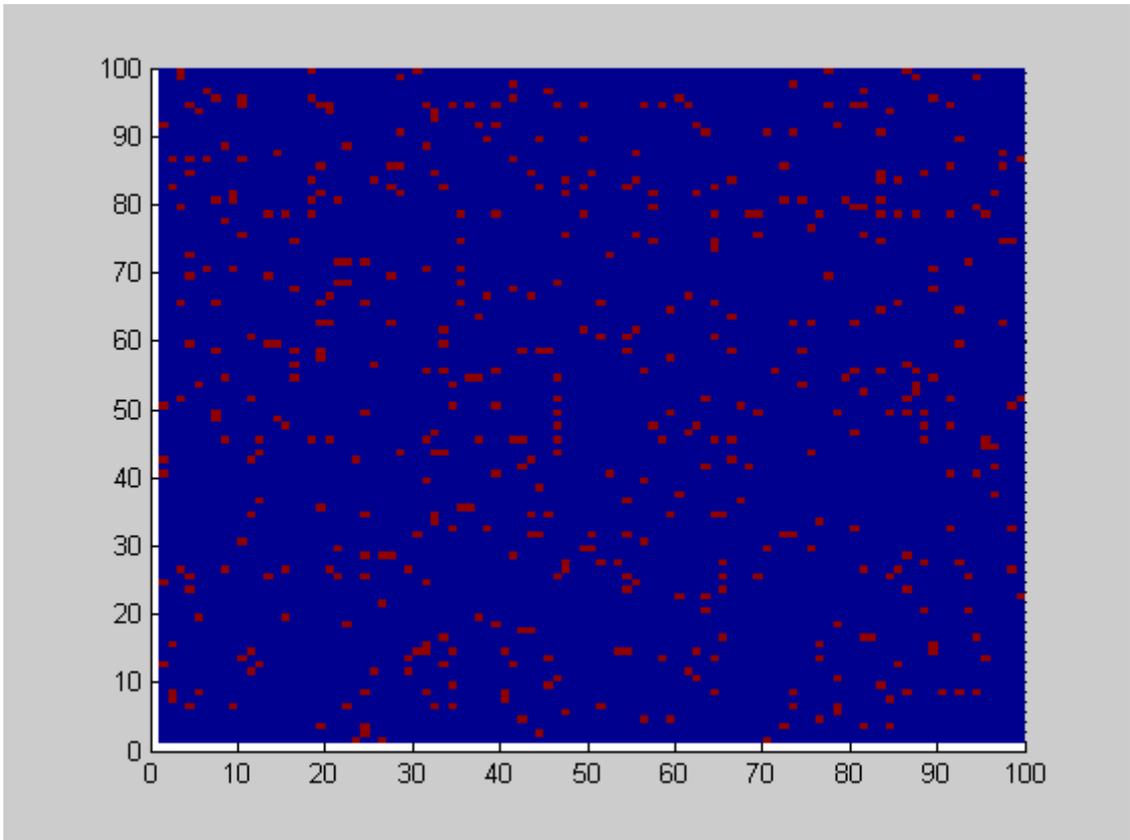

Figure 5a

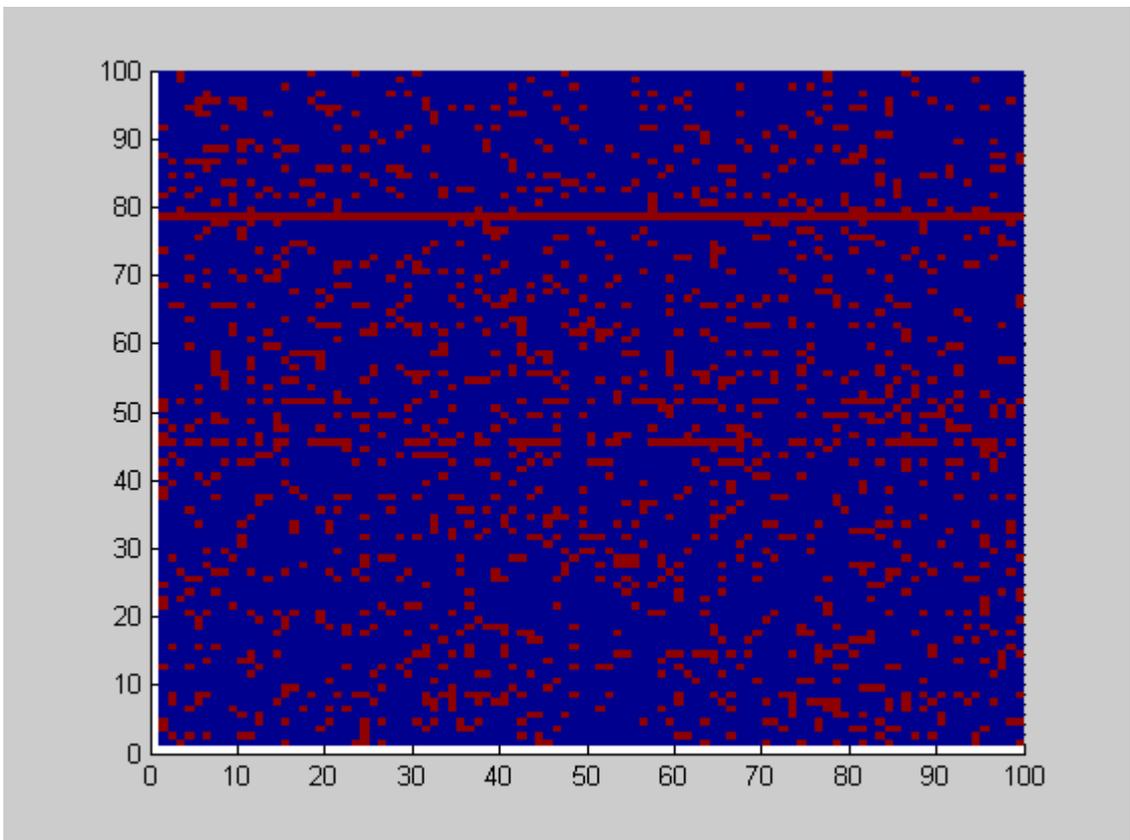

Figure 5b

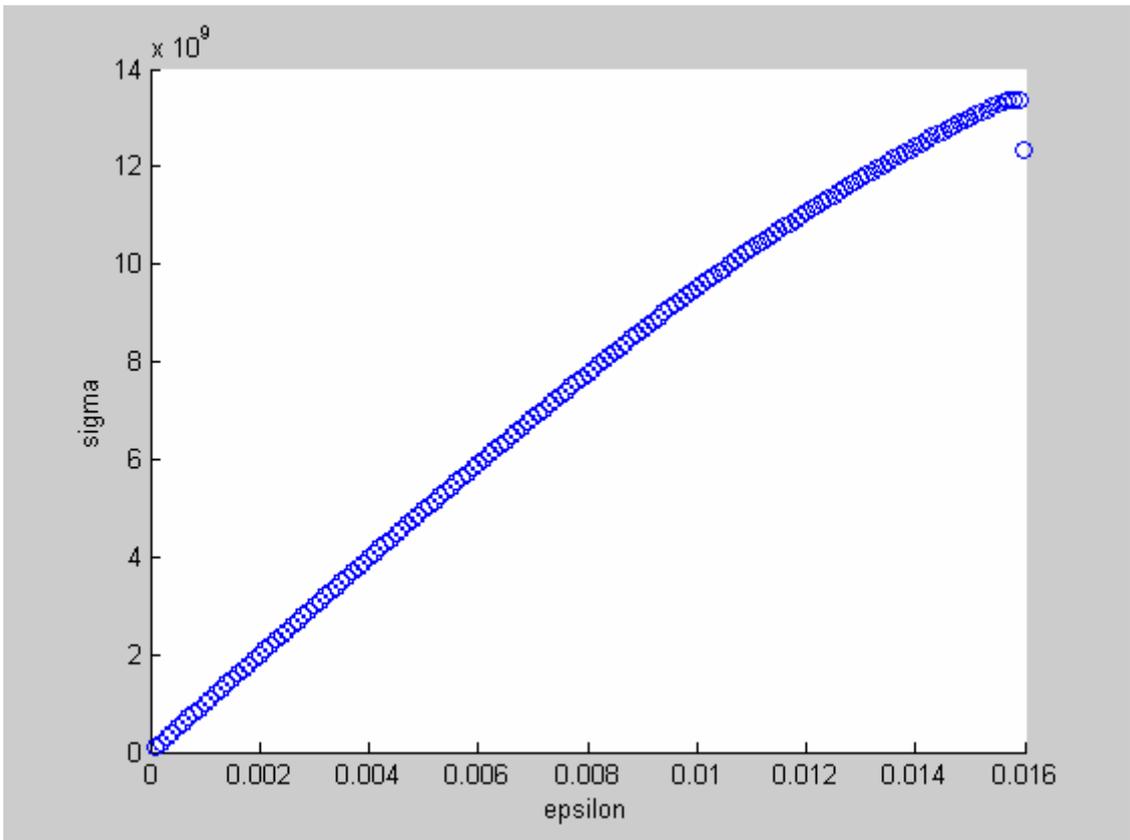

Figure 5c

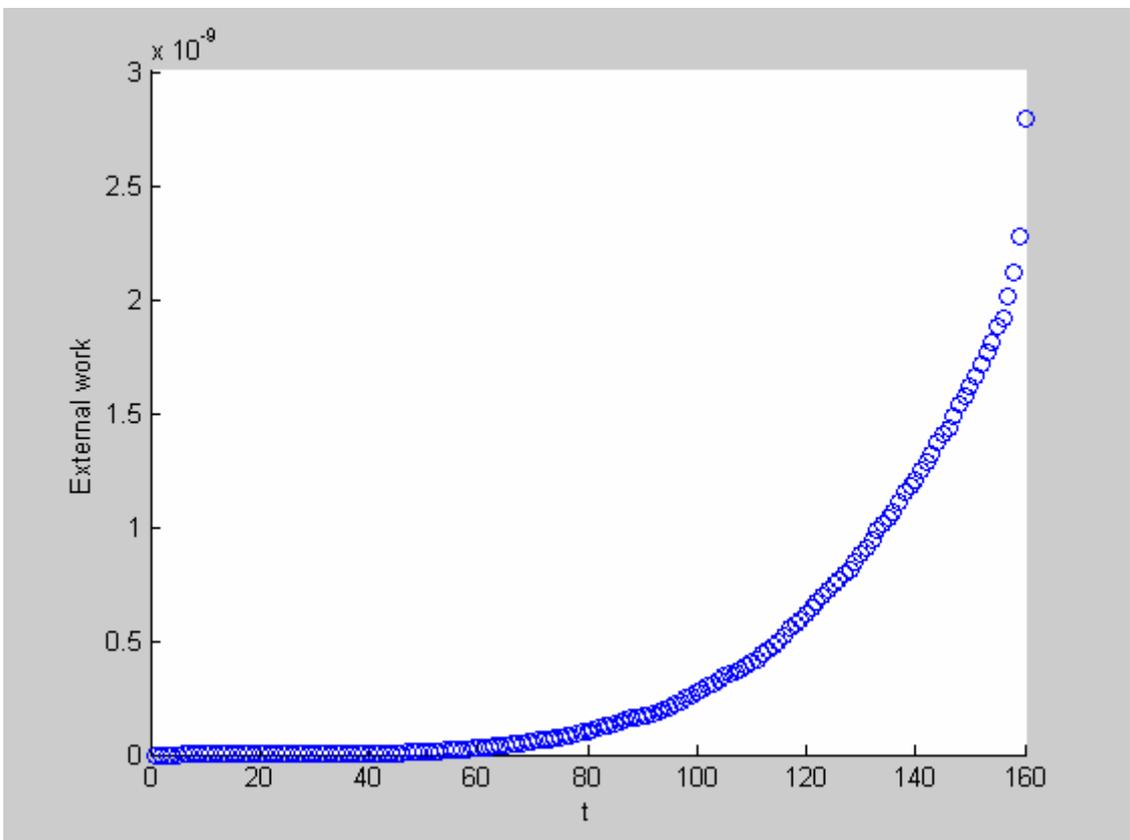

Figure 5d

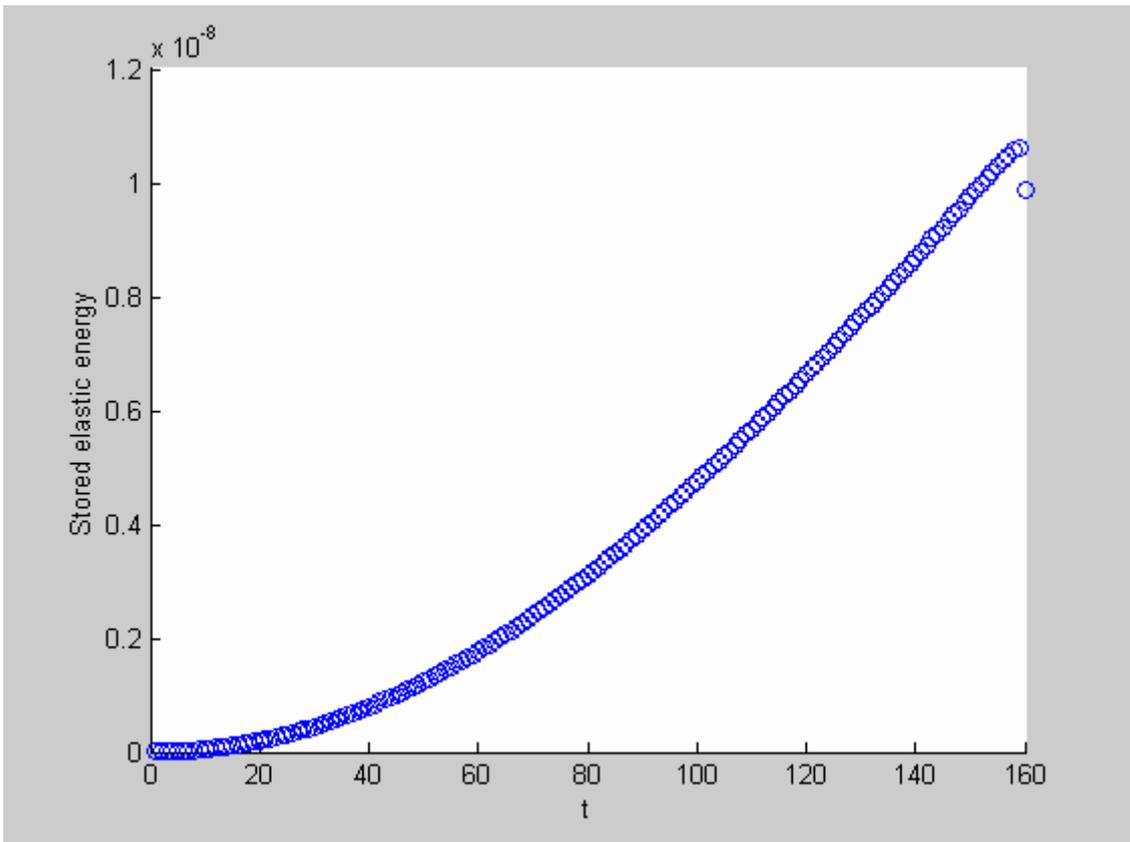

Figure 5e

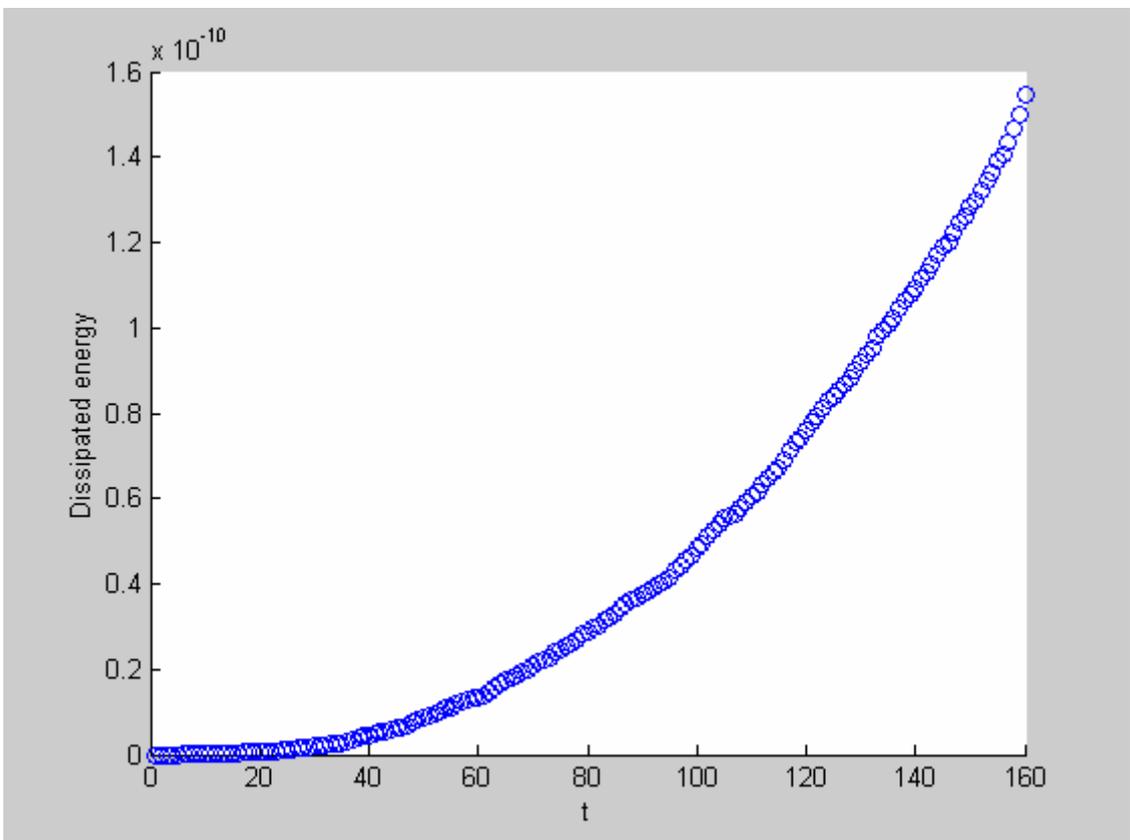

Figure 5f

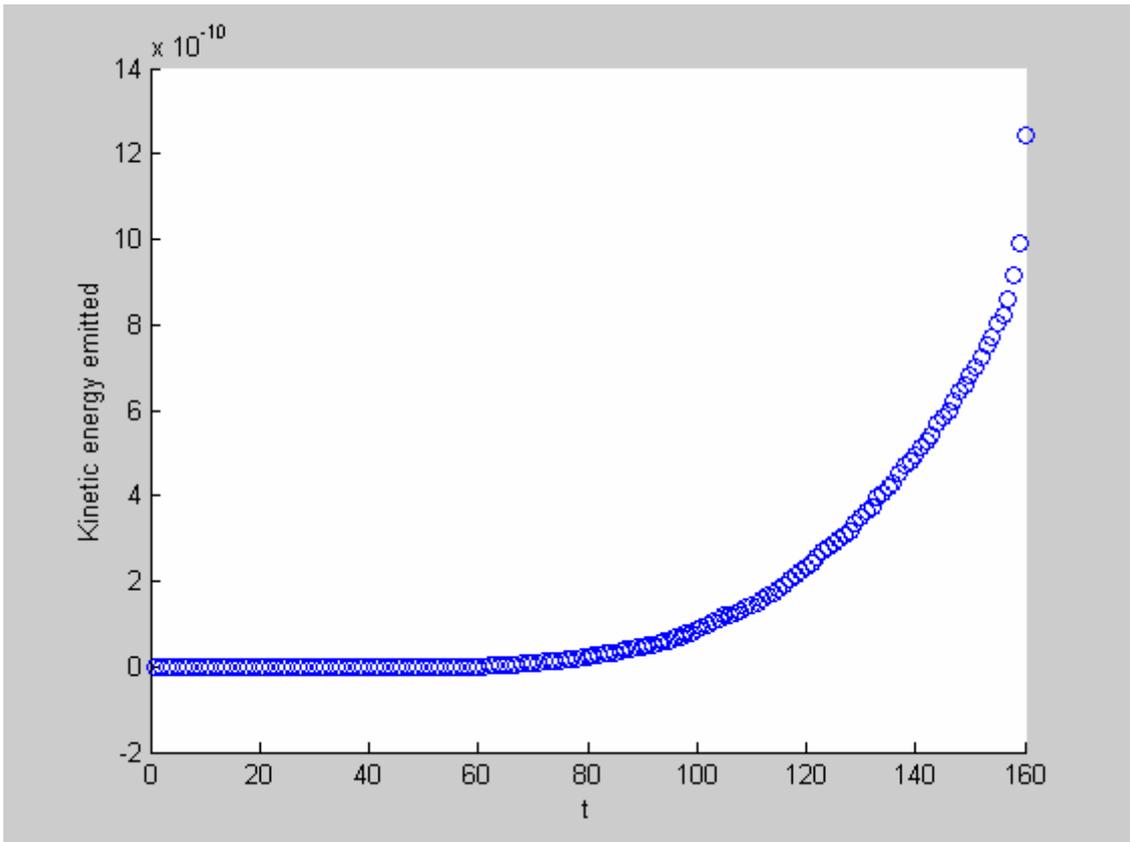

Figure 5g

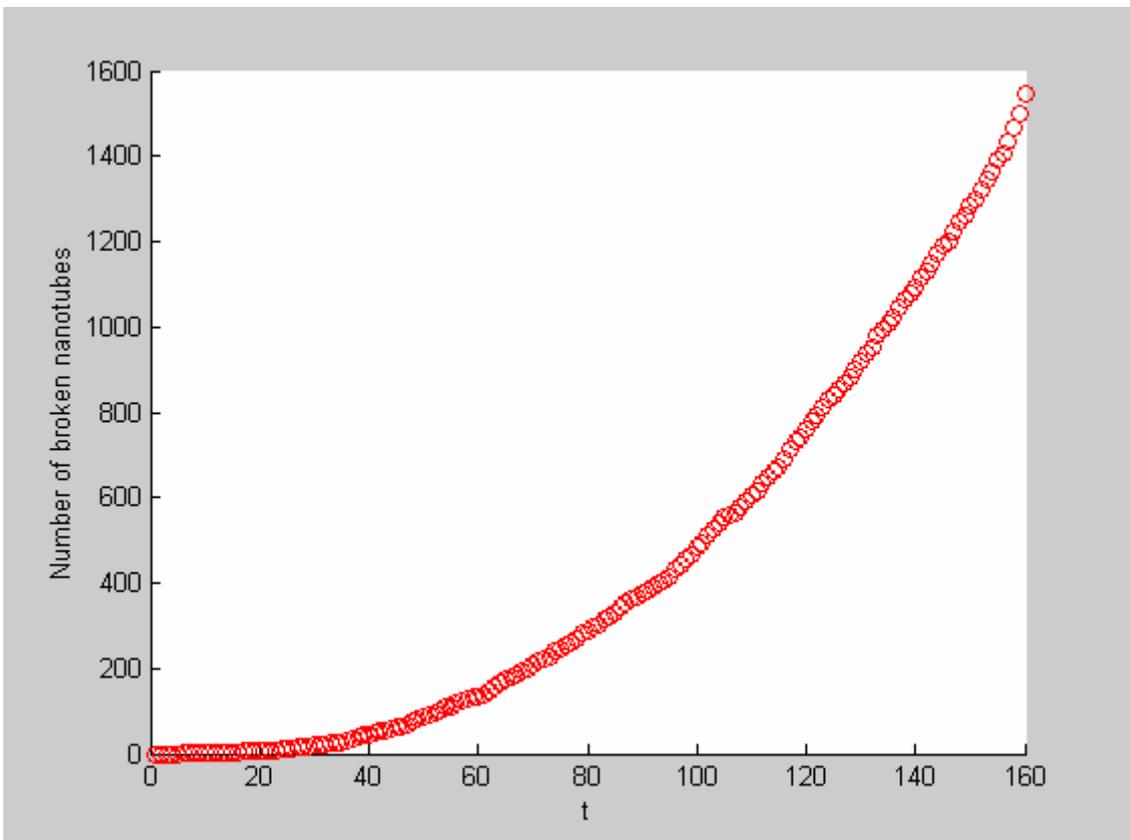

Figure 5h